%% file: ms.tex
\documentclass[preprint2]{aastex}
\shorttitle{Polarimetry of NGC 2264 Region}
\shortauthors{Kwon et al.}

\usepackage{times}
\slugcomment{To appear in the Astrophysical Journal Letters}

\begin{document}

\fontsize{10}{10.6}\selectfont
\title{NEAR INFRARED CIRCULAR POLARIZATION IMAGES OF NGC 6334-V}

\author{\sc Jungmi Kwon\altaffilmark{1,2}, Motohide Tamura\altaffilmark{1,2}, Phil W. Lucas\altaffilmark{3}, Jun Hashimoto\altaffilmark{1}, Nobuhiko Kusakabe\altaffilmark{1}, Ryo Kandori\altaffilmark{1}, Yasushi Nakajima\altaffilmark{4}, Takahiro Nagayama\altaffilmark{5}, Tetsuya Nagata\altaffilmark{6}, and James H. Hough\altaffilmark{3}  }
\affil{1. National Astronomical Observatory of Japan, 2-21-1 Osawa, Mitaka, Tokyo 181-8588, Japan}
\affil{2. Department of Astronomical Science, Graduate University for Advanced Studies (Sokendai), 2-21-1 Osawa, Mitaka, Tokyo 181-8588, Japan }
\affil{3. Centre for Astrophysics Research, University of Hertfordshire, College Lane, Hatfield AL10 9AB, UK}
\affil{4. Center of Information and Communication Technology, Hitotsubashi University, 2-1 Naka, Kunitachi, Tokyo 186-8601, Japan}
\affil{5. Department of Astrophysics, Nagoya University, Nagoya 464-8602, Japan}
\affil{6. Department of Astronomy, Kyoto University, Kyoto 606-8502, Japan}

\begin{abstract}

\fontsize{10}{10.6}\selectfont

We present results from deep imaging linear and circular polarimetry of the massive star-forming region NGC 6334-V. 
These observations show high degrees of circular polarization (CP),
as much as 22 \% in the $K_s$ band, in the infrared nebula associated with the outflow.
The CP has an asymmetric positive/negative pattern and is very extended ($\sim 80\arcsec$ or 0.65 pc). 
Both the high CP and its extended size are larger than those seen in the Orion CP region. 
Three-dimensional Monte Carlo light-scattering models are used to show that 
the high CP may be produced by scattering from the infrared nebula followed by
dichroic extinction by an optically thick foreground cloud containing aligned dust grains.
Our results show not only the magnetic field orientation of around young stellar objects but also the structure of circumstellar matter 
such as outflow regions and their parent molecular cloud along the line of sight.
The detection of the large and extended CP in this source and the Orion nebula may imply the CP origin of the biological homochirality on Earth.

\end{abstract}

\keywords{circumstellar matter -- dust, extinction -- infrared: stars -- ISM: individual objects (NGC 6334) -- polarization -- stars: formation}

\section{INTRODUCTION}

Magnetic fields have been thought for many years to play a crucial role in regulating accretion onto protostars, 
both in powering and shaping outflows and removing angular momentum from disk material, thereby allowing the protostar to gain mass
(Shu et al. 1987; Bergin \& Tafalla 2007; McKee \& Ostriker 2007). 
However, the precise role of the magnetic field is poorly understood and 
evidences for the shape and structure of the magnetic fields near the outflow regions 
have been difficult to obtain, although polarimetry is a technique that can certainly help. 

Imaging linear polarimetry has been used widely in recent years to investigate the environments of young stellar objects (YSOs). 
The measurements can be used to place constraints on the scattering geometries and on the scattering properties 
of the dust particles (e.g. Chrysostomou et al. 1996; Tamura et al. 2006), 
as well as magnetic field geometries (e.g. Chrysostomou et al. 2007). 

Linear polarization (LP) is easily produced and widely observed in star-forming regions 
being produced by scattering from dust (resulting in reflection nebulae), and dichroic extinction of starlight by aligned grains.
On the other hand, circular polarization (CP) is much less observed in star-forming regions.
There are only five star formation regions with near-infrared (near-IR) circular polarimetric studies published in refereed journals:
OMC 1, HH 135--136, R CrA, Cha IRN, and GSS 30 (see Clayton et al. 2005). 
We consider that the origins of CP may be produced by (1) multiple scattering from spherical grains,
(2) single scattering from aligned non-spherical grains,
(3) dichroic extinction with a systematic change in grain orientation,
and (4) dichroic extinction of scattered light.
Young bipolar nebulae that are seen in polarized scattered light at near-IR wavelengths are those rather rare exceptions 
among deeply embedded YSOs driving bipolar outflows. 
Circular polarimetry is therefore important in the study of stellar environments,  
giving information on scattering processes in the denser regions, on dust properties and magnetic field geometries.

The IR source NGC 6334-V consists of a double-lobed bright reflection nebula seen against a dark region, 
probably an optically thick molecular cloud, 
and was first observed by McBreen et al. (1979) at $\lambda \sim 70 \mu$m. 
The distance was estimated as $\sim 1.74 \pm 0.31$ kpc (Neckel 1978) and its total luminosity was measured to be $\sim 10^5 L_{\sun}$ 
(Harvey \& Gatley 1983; Loughran et al. 1986; McBreen et al. 1979). 
There are many linear polarimetric studies for this region 
(Simon et al. 1985; Nakagawa et al. 1990; Chrysostomou et al. 1994; Hashimoto et al. 2007, 2008; Simpson et al. 2009). 
Simpson et al. (2009) combined high-resolution {\it Hubble Space Telescope}/NICMOS data with {\it Spitzer}/GLIMPSE and radio data 
to show that the illuminating source of the nebula is a mid-IR and radio source 
located $\sim 2\arcsec$ south of the line connecting the two lobes. 
M\'{e}nard et al. (2000) reported in a conference proceedings that the NGC 6334-V has a relatively high CP ($\sim$23\%) 
but its details were not presented.
In this Letter, we present deep and wide near-IR LP and  CP images of the NGC 6334-V region.

\section{OBSERVATIONS AND DATA REDUCTION}

Observations were carried out using the SIRPOL imaging polarimeter 
on the Infrared Survey Facility (IRSF) 1.4 m telescope at the South African Astronomical Observatory. 
SIRPOL consists of a single-beam polarimeter and an imaging camera (Nagayama et al. 2003). 
The camera, SIRIUS, has three 1024 $\times$ 1024 HgCdTe IR detector arrays. 
SIRPOL enables wide-field (7\farcm7 $\times$ 7\farcm7 with a scale of 0\farcs45 pixel$^{-1}$) imaging polarimetry 
in the $J$, $H$, and $K_s$ bands simultaneously (Kandori et al. 2006).
The polarimetry unit consists of a stepped achromatic half-wave (for linear polarimetry)
or achromatic quarter-wave (for circular polarimetry) plate followed by a high-efficiency polarizer.

The linear polarimetry observations were made in 2006 July. 
The total integration time was 900 s per wave plate angle, 
and the stellar seeing size during the observations was $\sim$1\farcs2 in the $K_s$ band. 
We performed 10 s exposures at four wave plate angles (in the sequence 0\arcdeg, 45\arcdeg, 22\fdg5, and 67\fdg5) 
at 10 dithered positions for each set to measure the Stokes parameters. 
The same observation sets were repeated nine times toward the target object, NGC 6334-V, 
and for the sky background to increase the signal-to-noise ratio. 

The data were processed using IRAF in the standard manner, 
which included dark-field subtraction, flat-field correction, sky-bias subtraction, and frame registration. 
Further processing of data such as removing artificial stripe patterns or performing registrations was performed as described by Kwon et al. (2011).
Polarimetry of extended sources was carried out on the combined intensity (Stokes $I$) images  for each exposure cycle 
(a set of exposures at four wave plate angles at the same dithered position). 
The Stokes parameters $I$, $Q$, and $U$ were given by
\begin{equation}
   I = {1 \over 2} (I_{0} + I_{22.5} + I_{45} + I_{67.5}),
\end{equation}
\begin{equation}
   Q = I_{0} - I_{45},
\end{equation}
and
\begin{equation}
   U = I_{22.5} - I_{67.5},
\end{equation}
where $I_a$ is the intensity with the half wave plate oriented at $a\arcdeg$.
The linear polarization degree $P_{\rm linear}$ and the polarization position angle $\theta$ were  given  by
\begin{equation}
P_{\rm linear} = {{\sqrt{Q^2 + U^2}} \over {I}}       
\end{equation}
and
\begin{equation}
  \theta = {1\over2} \arctan {U \over Q}.
\end{equation}
The degree of polarization $P_{\rm linear}$ was corrected using the polarization efficiencies of SIRPOL,
which are 95.5\%, 96.3\%, and 98.5\% in the $J$, $H$, and $K_s$ bands, respectively.
We detected not only extended sources but also a number of point-like sources in the field of view, 
but the polarimetric studies of these point-like sources will be presented elsewhere. 

The circular polarimetry observations were made in 2007 June. 
For these the polarimeter unit attached to SIRIUS had a stepped achromatic quarter-wave plate.
An achromatic half-wave plate was continuously rotated above the quarter-wave plate 
thereby smearing out any LP present which might be measured as CP.
The instrumental CP was confirmed to be less than $\sim 0.3$\% with $\sim 100$\% incident LP.
The total integration time per two wave plate angles was 900 s (in the sequence of 0$\arcdeg$ and 90$\arcdeg$), 
and the stellar seeing size during the observations was $\sim 1\farcs4$ in the $K_s$ band. 
The uncertainty of polarization measurements due to the stability of the sky was reported  to be about 0.3\% by Kandori et al. (2006).

The CP images were obtained using
\begin{equation}
   V = I_{0} - I_{90}
\end{equation}
and
\begin{equation}
   I = I_{0} + I_{90}.
\end{equation}
The CP degree images are then derived using 
\begin{equation}
P_{\rm circular}    = V/I,
\end{equation}
and the degree of polarization $P_{\rm circular}$ was corrected
using the polarization efficiencies of SIRPOL.

\section{RESULTS}

\input{fig01.tex}

\input{fig02.tex}

Figure 1 shows a $J$-$H$-$K_s$ color composite Stokes $I$ image of the entire NGC 6334-V region. 
The bipolar nebula at the center of Figure 1, having considerably more significant CP,  
is mainly discussed in this Letter (hereafter NGC 6334-V IRN). 

As mentioned above, the NGC 6334-V IRN consists of a double-lobed bright reflection nebula, although it is asymmetric; 
with the brighter eastern (bluer) nebula larger and more extensive than the western (redder) nebula,
although the latter is somewhat intermittently extended further to the west. 
In Figure 2, the maximum extent of the eastern nebula is $\sim 30\arcsec$ (0.24 pc), 
while that of the western nebula is about 40$\arcsec$ (0.33 pc), including the intermittent extension. 

We have detected two CP regions in the NGC 6334-V field.
One is seen in the NGC 6334-V IRN,
while the other is in a small nebula in the northeast part of Figure 1 even though it is very weak and faint. 
The NGC 6334-V IRN shows very large CP region, 
and its whole size including an additional faint region with high negative CP ($-11$\%) 
is $\sim 80\arcsec$ (0.65 pc).
It is larger than the CP region of OMC-1 (210$\arcsec$ equivalent to 0.45 pc at 450 pc, see Figure 2(e)). 
The CP of the NGC 6334-V IRN shows a quadrupolar structure of positive and negative degrees of CP, 
even though there is an obvious enhancement of negative CP that is spatially coincident with the region of the western nebula. 
The absence of a symmetric CP pattern (Hough et al. 2001) shows that the NGC 6334-V IRN region is relatively complex.

\section{DISCUSSION AND MODELS OF THE SCATTERING GEOMETRY}

\subsection{Model Configuration}

The western nebula, which has high CP, appears very red in the near-IR color image, 
whereas the much bluer eastern nebula has much lower CP. 
This suggests that the observed CP is related to the amount of extinction by aligned grains and scattering, 
as has been found in previous studies
(see OMC-1; Bailey et al. 1998; Buschermohle et al. 2005; Fukue et al. 2009, or HH 135-136; Chrysostomou et al. 2007). 

Furthermore, previous simulations (Lucas et al. 2005) have indicated that 
dichroic scattering cannot produce CP $> 20$\% 
(and simultaneously produce the observed high LP and non-negligible albedo) without employing extreme grain axis ratios ($>$3:1). 
Similarly, several authors have shown that multiple scattering from spherical grains can only produce low CP 
(e.g., Fischer et al. 1996).

We have therefore constructed a model of the western nebula 
in which dichroic extinction of linearly polarized scattered light is the main mechanism producing the observed CP. 
We used the $shadow.f$ light-scattering code (Lucas et al. 2004) to simulate the data.
The model reproduces several features of the data and makes some attempt to match the actual structure of the western nebula.
Since it is not possible to reliably model the detailed three dimensional physical structure 
of the nebula with only two dimensional data as a constraint, 
it should be regarded as an illustrative model rather than an exact match.

\input{fig03.tex}
\input{fig04.tex}

Figure 3 illustrates the modeled geometry of the NGC 6334-V IRN and 
Figure 4 presents the results of Monte Carlo simulations using the geometry shown in Figure 3. 
It includes a large optically thick cloud in front of the NGC6334-V IRN, 
as suggested by Simpson et al. (2009) on the basis of the large dark region seen in GLIMPSE data and in our $JHK_s$ images.
In the plane of the sky we have the illuminating protostar surrounded by a disk and a small envelope, 
and two separate clouds to the west that represent the western nebula. 
The smaller of these two clouds was included because it represents the region of highest CP (22\%, Figure 2). 
We note that the protostar and the disk and envelope are not observed in the near-IR, 
so we do not place constraints on their properties.
However, the constraints on many parameters such as the grain axis ratio and optical depth in each cloud are reasonably tight,
as will be shown in Paper II.

\subsection{Simulation of the $B$ Field }

We assume in the model that the magnetic field is uniform, being in the plane of the sky and oriented 27$\arcdeg$ east of north. 
This field orientation reproduces the sense of positive and negative Stokes $V$ observed in the western nebula. 
We note that this approximately north--south field orientation is consistent with the position angle of the LP vectors for stars 
near NGC 6334-V IRN in Figure 2, though they show some scatter.
If we have an east--west field then the sign of CP is reversed. 
It is important to note that the CP in the model is produced almost entirely by dichroic extinction in the foreground cloud, 
so the model does not allow us to determine the orientation of the field in the two clouds or in the immediate vicinity of the protostar. 

Using our SIRPOL data for the region we estimated $A_v \sim 37$ mag ($A_k \sim 4.2$) 
in the same manner as in Tamura et al. (1997),
i.e. optical depth, $\tau_K \sim$ 3.9 for the western nebula. 
We note that this is somewhat higher than the extinction $A_v \sim 13$ mag ($A_k \sim 1.5$) 
measured in the HH 135--136 system by Tamura et al. (1997). 
This is almost certainly the reason why the maximum CP was only 8 \% in that system (Chrysostomou et al. 2007), 
compared with 22 \% in NGC 6334-V.
We adopt $\tau_K = 3.9$ as the line-of-sight optical depth for the foreground cloud in the model. 
By contrast, the two clouds that represent the model western nebula have optical depths, $\tau_K < 1$. 
This is to be consistent with the observation that the surface brightness of the western nebula shows no obvious decline 
with increasing distance from the illuminating protostar (see Figure 2, at R.A. offsets 0$\arcsec$ to $-$20\arcsec), 
which would be the case if there were substantial extinction within the western nebula itself. 

The observed maximum CP of 22 \% is reproduced with oblate dust grains with an axis ratio of 1.025:1, 
and a size distribution described by $a_{\rm min}$ = 0.005 $\mu$m, $a_{\rm max}$ = 0.35 $\mu$m, and an $a^{-3.5}$ power law. 
This is similar to the axis ratio of 1.03:1 that was found for the HH 135--136 system in Chrysostomou et al. (2007). 
The grain albedo is 0.27 at 2.2 $\mu$m. 
We note that a similar model with $a_{\rm max}$ = 0.75 $\mu$m led to a very obvious effect of dichroic extinction on the LP pattern 
(aligned vectors, which are not observed) and greatly reduced the percentage of LP. 
The grain composition is a mixture of silicate and small amorphous carbon grains, taken from Lucas et al. (2005). 
As noted therein, the amorphous carbon serves to provide an absorptive component. 
This provides a mixture with plausible optical properties but is not intended to be taken as a chemically realistic model.
The $shadow.f$ code assumes that the grains are perfectly aligned such that the short axis 
of the oblate grains is paralleled to  the direction of the magnetic field. 

The model reproduces the observed degree of LP fairly well (60\%--90\% in the model, compared with 60\%--70\% observed) 
as well as the general trend of north--south LP vectors observed in most of the region caused by dichroic extinction 
(Figure 2(c); see also Hashimoto et al. 2008).
The highest model LP and CP occurs in Cloud 2. 
This is due to the fact that Cloud 2 is located in the plane of the sky, so that the scattering angle is 90\arcdeg. 
By contrast, Cloud 1 is centered slightly in front of the plane of the sky, so that the scattering angle is smaller. 
This gives lower LP in Cloud 1 and lower CP also, since CP is proportional to LP 
(specifically the Stokes $U/I$ component of LP, for a Stokes $Q$ axis defined by the magnetic field direction). 
A feature of the model is that Cloud 1 is tilted in the plane of the sky (see Figure 3) 
so that the region of negative Stokes $V$ is closer to the protostar and the region of positive Stokes $V$ is further away, 
as is seen in the data. 

In an attempt to reproduce the gradients in LP and CP that are observed, Cloud 1 is also tilted in the plane 
containing the line of sight ($XY$ plane) such that the region of positive Stokes $V$ is closest to the plane of the sky 
(giving a scattering angle near 90\arcdeg) and the region of negative $V$ is further in front. 
This causes the region of positive $V$ to have a higher LP (up to 80\%) and higher CP ($+$21\%) than the region of negative $V$, 
which has LP $\sim 60$\% and CP $\sim -19$\%. 
In fact the regions of positive and negative $V$ in the western nebula show greater differences in the percentage of CP: 
$+$16\% and $-$11\% respectively, so this aspect of the model is only partly successful. 
This illustrates the difficulties in trying to model the line of sight aspects of the nebula structure with two-dimensional image data.

\subsection{The Eastern Nebula}

We have not attempted to model the eastern nebula because the observed structure of 
negative $V$ closer to the protostar and positive $V$ further from the protostar cannot readily be explained 
if the illuminating source is at the location indicated in Figure 2.
Hashimoto et al. (2007) concluded that the eastern nebula is illuminated by a different YSO and 
Simpson et al.(2009) observed that there may indeed be a second mid-IR source $\sim 3\arcsec$ north of the protostar 
that illuminates the western nebula. Here we simply note that the eastern nebula has a much lower CP 
( $< 2$ \% over most of the area) so the other CP mechanisms mentioned in the Introduction may be contributing. 
This would further complicate any analysis of that region.

\section{CONCLUSIONS AND IMPLICATIONS}

Our results for the LP are very consistent with previous studies,
while the CP, which can reach $\sim 22$\%, provides the first detailed images for this region;
the CP pattern is asymmetric.
The degrees of CP and its distribution are larger than that of OMC-1.
Our model shows not only the magnetic field orientation of around YSOs but also the structure of circumstellar matter 
such as outflow regions and surrounding parent molecular clouds along the line of sight.

The high degree of circularly polarized light reported in high-mass star-forming regions is an attractive explanation
for the origin of homochirality on Earth (Bailey et al. 1998; Fukue et al. 2010),
given the strong evidence that the solar system originated in a high-mass star-forming region (Miller \& Scalo 1978).
An enantiomeric excess may have been produced, 
through asymmetric photolysis in prebiotic organic molecules present in the pre-solar nebula
and delivered to Earth during the heavy bombardment phase.

\acknowledgments
The authors thank the referee, Professor Masafumi Matsumura, for his careful reading of the manuscript and for his helpful suggestions to improve this Letter. 
J.K. is supported by the JSPS Research Fellowships for Young Scientists  (DC2: 24$\cdot$110).
M.T. is supported by the MEXT, Grants-in-Aid 19204018 and 22000005.

\clearpage

\end{document}

%% file: fig01.tex
\begin{figure*}
  \centering
    \includegraphics[width=0.9\textwidth]{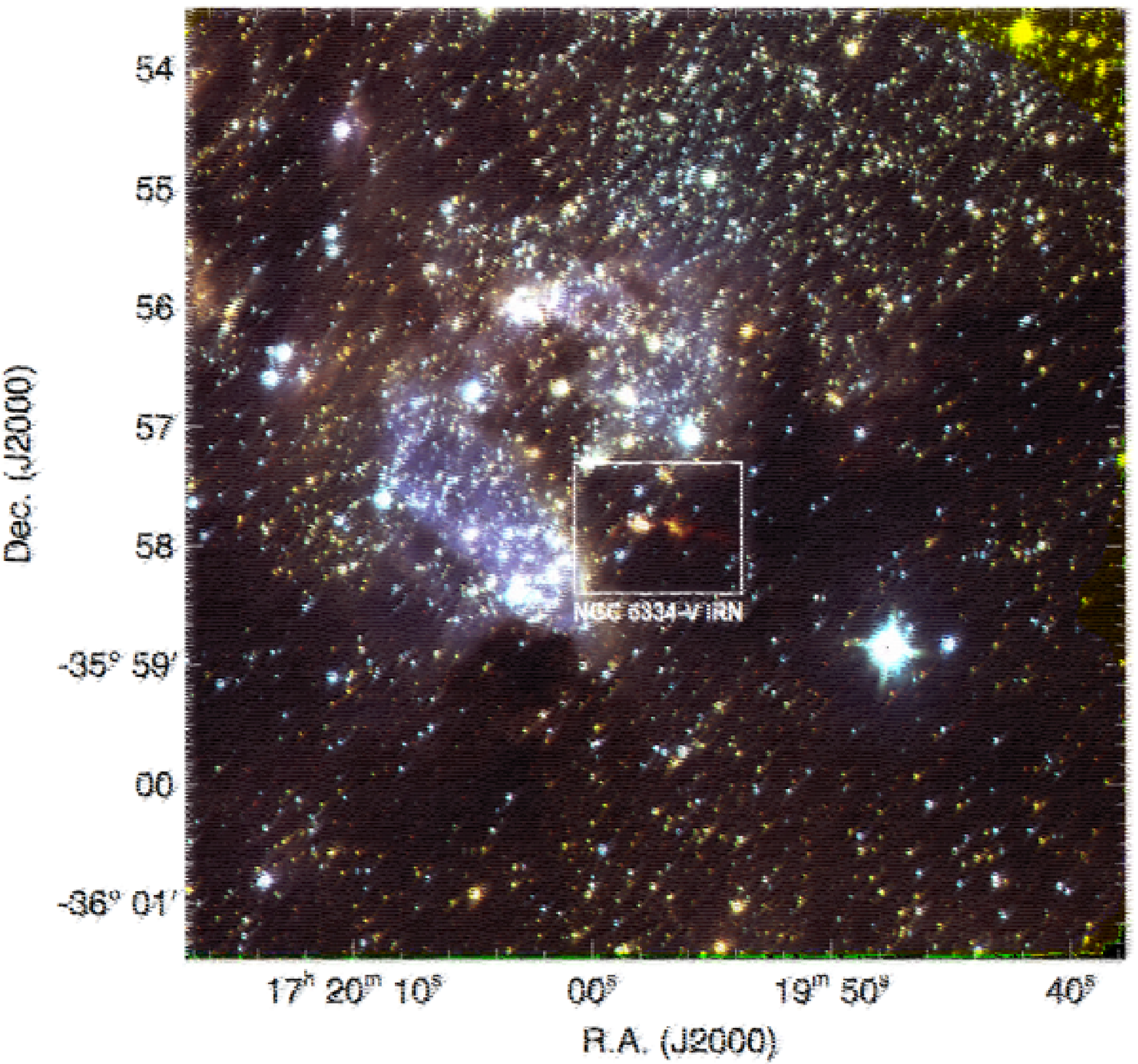}
\caption{
Color composite Stokes $I$ image of the NGC 6334-V region in the $J$ (blue), $H$ (green), and $K_s$ (red) bands 
from the IRSF/SIRPOL (CP) observations. 
The thin lines of red, green, and blue around the perimeter are boundaries of Stokes $I$ images associated with each band. 
Note that there are bad pixel clusters around the upper-left and upper-right corners and the middle of the right boundary. 
NGC 6334-V IRN is labeled.}
\end{figure*}

%% file: fig02.tex
\begin{figure*}
  \centering
    \includegraphics[width=0.7\textwidth]{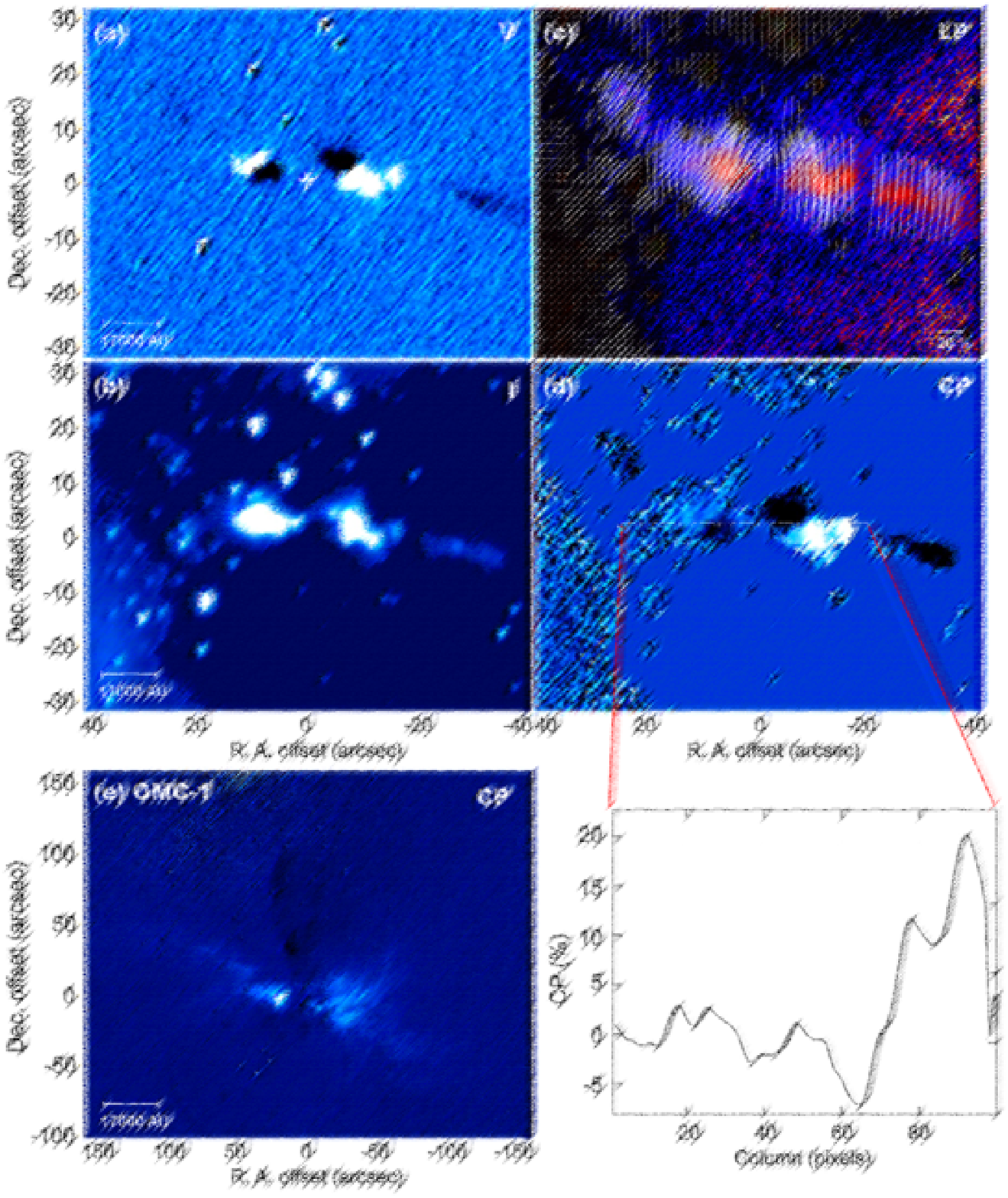}
\caption{
((a) and (b)) Stokes $V$ and $I$ images of NGC 6334-V IRN in the $K_s$ band, respectively. The white cross shown in (a) is the location of the illuminating star (Simpson et al. 2009). 
(c) $K_s$ polarization vector map of NGC 6334-V IRN superposed on the LP image.
(d) CP image in the $K_s$ band and a plot line indicated a white box and columns of the sub-CP image smoothed by 3 pixels of NGC 6334-V IRN.
For making the CP image, $I$ image was masked with a threshold of $2 \sigma$ of approximate sky value. 
(e) CP image of OMC-1 in the $K_s$ band. 
Note that bright point-like sources are not completely canceled and are visible in this image even if they are unpolarized, 
due to slight misalignment among different images. }
\end{figure*}

%% file: fig03.tex
\begin{figure*}
  \centering
    \includegraphics[width=0.42\textwidth]{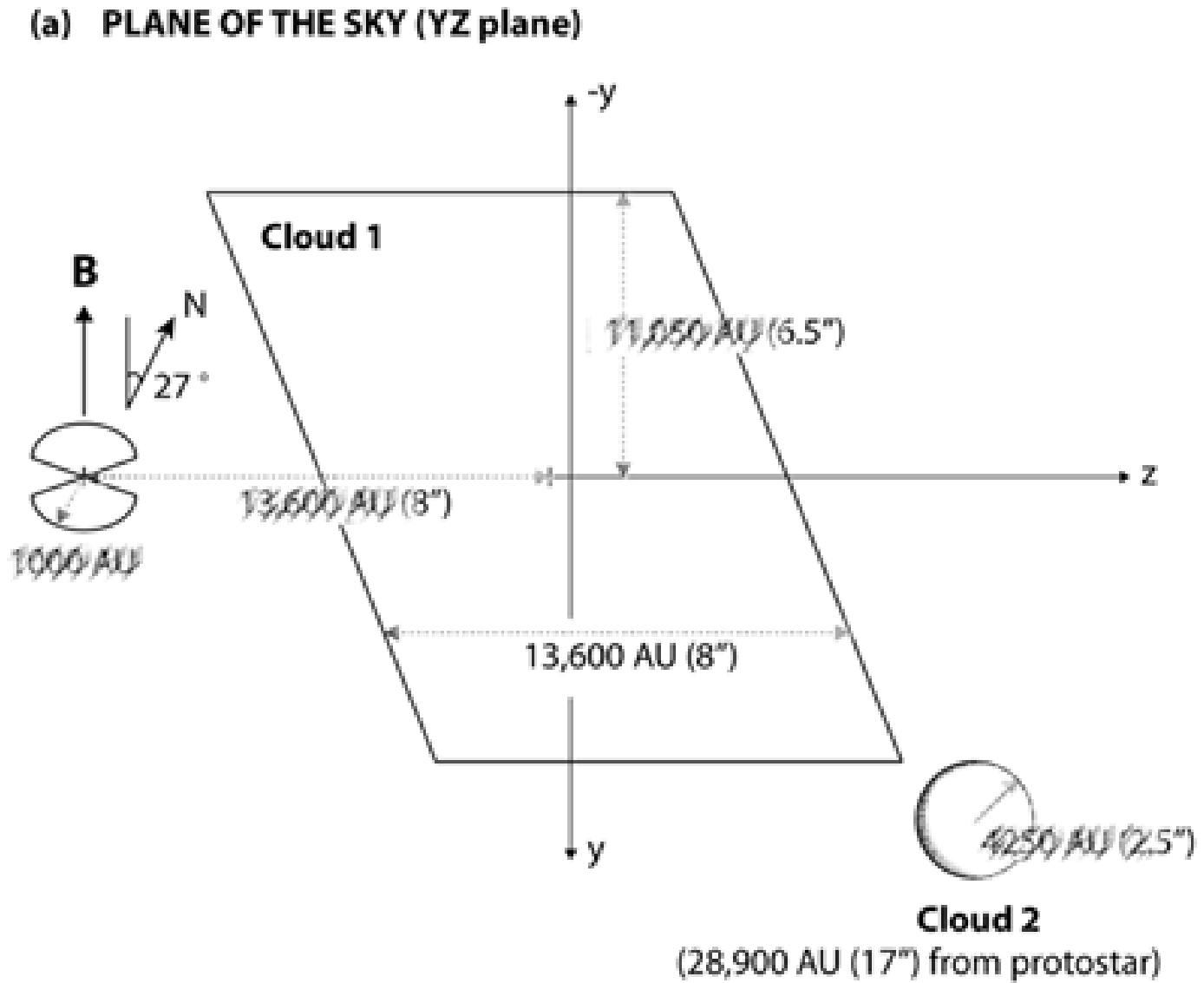}
    \includegraphics[width=0.35\textwidth]{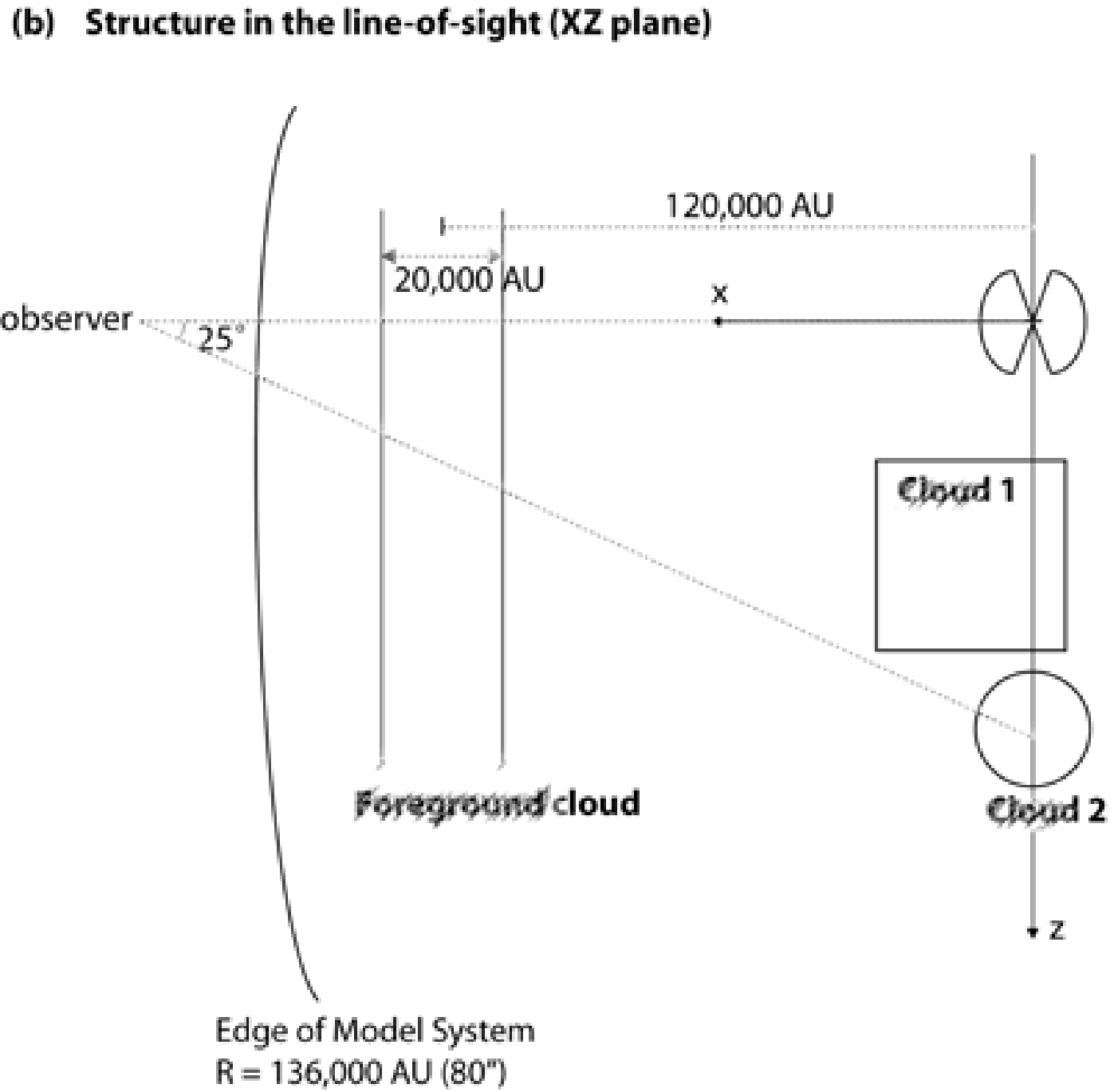}
\caption{
((a) and (b)) Stokes $V$ and $I$ images of NGC 6334-V IRN in the $K_s$ band, respectively. The white cross shown in (a) is the location of the illuminating star (Simpson et al. 2009). 
(c) $K_s$ polarization vector map of NGC 6334-V IRN superposed on the LP image.
(d) CP image in the $K_s$ band and a plot line indicated a white box and columns of the sub-CP image smoothed by 3 pixels of NGC 6334-V IRN.
For making the CP image, $I$ image was masked with a threshold of $2 \sigma$ of approximate sky value. 
(e) CP image of OMC-1 in the $K_s$ band. 
Note that bright point-like sources are not completely canceled and are visible in this image even if they are unpolarized, 
due to slight misalignment among different images. }
\end{figure*}

%% file: fig04.tex
\begin{figure*}
  \centering
    \includegraphics[width=0.63\textwidth]{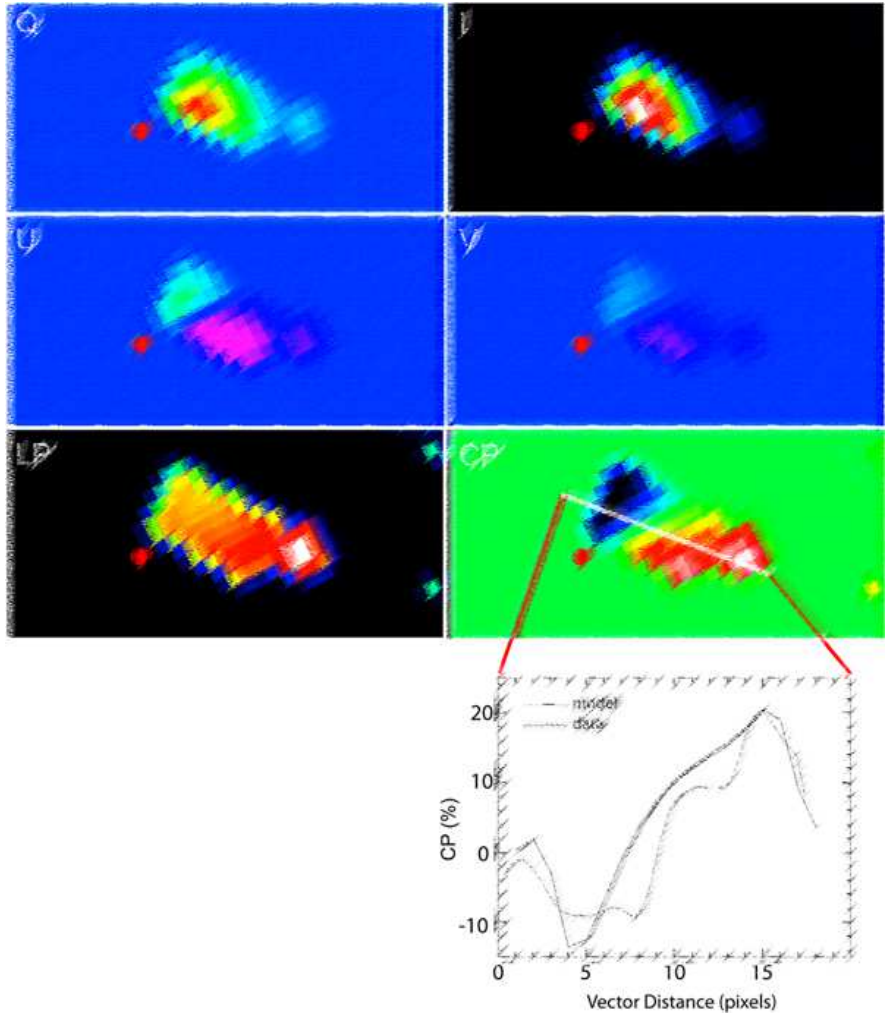}
\caption{
Results of Monte Carlo simulations with grain axis ratio of 1.025:1. The location of the protostar is indicated with a circle. 
A profile of CP in the model and in the data is also shown.
Note that the LP and CP images are masked at a level (Stokes $I$ $\leq$ 0.01). 
}
\end{figure*}